# LARGE-SCALE MULTIGRID
# WITH ADAPTIVE GALERKIN COARSENING

FABIAN BÖHM‡∗, NILS KOHL†, HARALD KÖSTLER∗‡, AND ULRICH RÜDE‡§

**Abstract.** We propose a robust, adaptive coarse-grid correction scheme for matrix-free geometric multigrid targeting PDEs with strongly varying coefficients. The method combines uniform geometric coarsening of the underlying grid with heterogeneous coarse-grid operators: Galerkin coarse grid approximation is applied locally in regions with large coefficient gradients, while lightweight, direct coarse grid approximation is used elsewhere. This selective application ensures that local Galerkin operators are computed and stored only where necessary, minimizing memory requirements while maintaining robust convergence. We demonstrate the method on a suite of sinker benchmark problems for the generalized Stokes equation, including grid-aligned and unaligned viscosity jumps, smoothly varying viscosity functions with large gradients, and different viscosity evaluation techniques. We analytically quantify the solver's memory consumption and demonstrate its efficiency by solving Stokes problems with $10^{10}$ degrees of freedom, viscosity jumps of $10^6$ magnitude, and more than 100,000 parallel processes.

**1. Introduction.** Matrix-free geometric multigrid methods are among the most powerful methods for solving linear systems arising from the discretization of elliptic partial differential equations. Their performance and extreme scalability to up to trillions ($10^{12}$) of unknowns excels through exploitation of the underlying grid structure [27, 19, 3, 24]. However, matrix-free compute kernels and the dependence on a structured grid hierarchy render the solution of problems with strongly varying coefficients challenging. Such issues must be resolved with an appropriate discretization of the continuous problem and with a suitable solver for the linear system.

Effective multigrid methods for problems with jumping or strongly varying coefficients must deliver a suitable coarse grid correction that can usually only be constructed with operator-dependent interpolation operators $P = P(A)$ and matching coarse grid operators $A_c$, typically constructed via Galerkin coarsening $A_c = P^T A P$ from the fine-grid operator $A$ [2, 33] (referred to as Galerkin coarse-grid approximation (GCA)). For constant or smooth coefficients and standard interpolation techniques, the coarse grid operators are typically identical or similar to those that arise from a rediscretization of the continuous problem on the coarse grid, (referred to as direct coarse-grid approximation (DCA)). Standard matrix-free compute kernels essentially employ DCA on-the-fly and therefore cannot properly include a fine-grid operator dependence. Algebraic multigrid methods [29, 30] are a powerful class of methods that automatically construct the interpolation and coarse-grid operators from the fine-grid operator. However, this mechanism requires storing the matrices to enable random access to matrix entries, and thus typically cannot exploit the grid structure to design fast compute kernels. Needless to say, those methods are not matrix-free and come

∗Erlangen National High Performance Computing Center (NHR@FAU), Erlangen, Germany.
†Dept. of Earth and Environmental Sciences, Ludwig-Maximilians-Universität München (LMU), Munich, Germany (nils.kohl@lmu.de).
‡Friedrich-Alexander-Universität Erlangen-Nürnberg (FAU), Erlangen, Germany ({fabian.boehm,harald.koestler,ulrich.ruede}@fau.de).
§Centre Européen de Recherche et de Formation Avancée en Calcul Scientifique (CERFACS), Toulouse, France. VSB - Technical University of Ostrava, Ostrava, Czechia.

with a performance penalty and severely increased memory consumption.

**1.1. Contribution.** In Section 2, we present an adaptive multigrid coarsening strategy designed for large-scale PDEs with strongly varying or discontinuous coefficients. The method employs an element-wise, distributed formulation of GCA in regions where the coefficient exhibits large gradients, without requiring assembly of a global sparse matrix. In regions with low coefficient variation, the algorithm defaults to the fast and memory-efficient direct coarse approximation (DCA). Figure 1 illustrates the concept. The resulting multigrid cycles remain as lightweight as those based on pure DCA, while providing significantly improved robustness with respect to coefficient variation.

We find that the presented adaptive Galerkin coarse-grid approximation (AGCA) reduces the memory consumption of coarser operators to negligible amounts for large-scale problems with commonly encountered jump shapes. The algorithm modifies only the coarse-grid correction, while the expensive relaxation and residual computation on the fine-grid remain entirely matrix-free. Conceptually, the method can be viewed as a matrix-free, geometric multigrid scheme that adaptively switches to AMG-style coarse-grid correction only where demanded by the coefficient. For parallel scalability, the local GCA matrices are assembled in an element-wise manner, keeping the construction embarrassingly parallel and enabling kernels to exploit the grid structure.

To study our method, we consider the generalized, variable-coefficient Stokes equation as a challenging demonstrator (Section 3). This choice is motivated through applications from the geosciences, particularly the simulation of Earth mantle convection, where the momentum and mass balance are modeled through a Stokes system, that due to strong radial and lateral viscosity variations is difficult to treat numerically [27, 4, 8].

We also observe, that the specific problem setup with respect to the coefficient has major impact on the effectiveness of the coarse-grid correction. To demonstrate this and to evaluate the robustness of our proposed scheme, we define a suite of test problems featuring strongly varying viscosity, commonly known as (multi)sinker problems (Section 3, Section 4). These problems cover a broad range of scenarios, including grid-aligned and -unaligned viscosity jumps, and continuous viscosity fields with steep gradients. We also analyze the effects of viscosity averaging and smoothing on the algebraic convergence.

Finally, we evaluate the potential memory savings of our scheme (Section 5) and assess the scalability of the resulting multigrid solver for large-scale Stokes problems with up to $10^{10}$ degrees of freedom (DoFs), viscosity contrasts of up to $10^6$, and executions on more than 100,000 parallel processes (Section 6).

**1.2. Related Work.** The ubiquity of parameter heterogeneities in engineering and scientific applications led to the development of a variety of methods to tackle strongly varying coefficients in partial differential equations. They include discretization techniques like adaptive mesh refinement and a posteriori error estimation [7, 34, 9], tailored domain decomposition methods [26, 12, 22], and immersed finite element methods [21]. Alternatively, appropriate modifications of the continuous problem, e.g., through coefficient averaging [11, 32] may help.

Even with an appropriate discretization, in practice, especially for strong variations or discontinuities, the resulting linear systems often prove difficult to solve and standard solvers may converge slowly or diverge. Algebraic multigrid methods [29, 30] are known to effectively treat coefficient variations through the selection of operator-dependent interpolation operators and Galerkin coarsening. While the

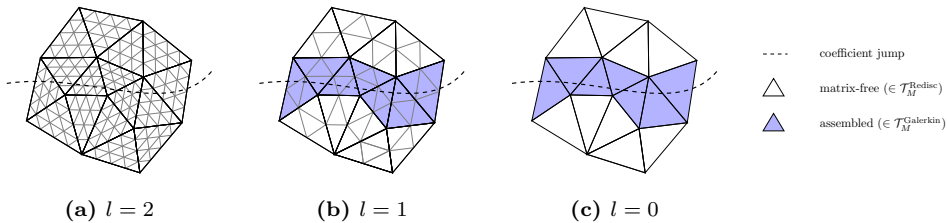

**(a)** $l = 2$     **(b)** $l = 1$     **(c)** $l = 0$

*Figure 1  Adaptive Galerkin coarse approximation (AGCA) in 2D.* The three figures show a multigrid hierarchy with three levels, the finest grid being $L = 2$. The curved dashed line indicates a jump (or large gradient) of a PDE coefficient. With the proposed adaptive Galerkin coarsening scheme, the assembly of the local coarse-grid element matrices is only executed on the colored elements. Those are exactly those elements that are not on the finest grid, and contained in macro-elements that intersect the jump. All non-colored elements are treated matrix-free, via direct coarse-grid approximation. Most importantly, the entire fine grid is treated fully matrix-free.

convergence rates of AMG are generally robust even in complicated setups, their computational performance, memory efficiency, and scalability cannot match that of optimized matrix-free geometric multigrid methods [8]. Still, impressive results and advances have been demonstrated especially for parallel AMG [1, 20, 10, 15]. AMG is popular in a hybrid setting for the solution of coarse grid problems when meshes in (matrix-free) geometric multigrid cannot be effectively coarsened further in parallel settings [31, 27, 18, 9]. [25] achieves improved performance by exploiting structure in a semi-structured AMG variant.

Specifically related to the coarse-grid correction, [16] effectively reduces communication costs and the nonzero density of AMG coarse-grid operators by truncating the operator stencils on coarser grids. Another interesting approach is presented in [35], where similar to the idea presented in this paper, a combination of geometric multigrid with algebraic coarse grid correction is designed. The authors compute, compress, and store the differences of the matrix entries that result from re-discretization and algebraic operator-dependent coarsening. In parts of the domain where the coefficients are smooth, the differences are negligible. While this approach is technically not completely matrix-free, it still can result in a similar memory-overhead compared to the technique presented in this paper.

**2. Adaptive Galerkin Coarse-grid Approximation (AGCA).** We aim to solve large, sparse linear systems of the form

$$(2.1) \qquad A\mathbf{x} = \mathbf{b}, \quad A \in \mathbb{R}^{N \times N},$$

which arise from the finite element (FE) discretization of partial differential equations (PDEs) with strongly varying or discontinuous coefficients. Since the proposed method is not limited to a specific PDE, the algorithms are formulated in a PDE-agnostic manner in this section. We consider a concrete example in Section 3.

Although the method does not require a specific grid type, the presence of a block structure simplifies its implementation. In our case, the finite element (FE)-discretization is defined on a tetrahedral, block-structured grid as described in [23]. First, the domain $\Omega$ is subdivided into an unstructured tetrahedral grid $\mathcal{T}_M$ consisting of unrefined *macro elements* $M$. We then apply $l$ uniform refinement steps to each macro element, denoted by $M_l$. The refinement process is depicted in Figure 2. The

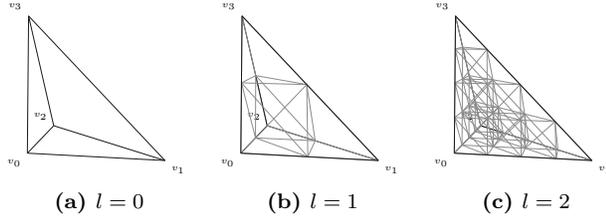

**(a)** $l = 0$     **(b)** $l = 1$     **(c)** $l = 2$

***Figure 2*** ***Uniform tetrahedral refinement.*** *The figure shows a single macro-element $M \in \mathcal{T}_M$ of the unstructured macro-grid $\mathcal{T}_M$ being refined two times as described in [23], yielding $M_1$ and $M_2$.*

child tetrahedra resulting from $l$ refinements of a macro element are called *micro elements* $m_l$, and we define the set of all micro elements as $\mathcal{T}_{m,l}$. Then, a block-structured tetrahedral grid is described by the following collection:

$$\Omega = \bigcup_{M \in \mathcal{T}_M} M = \bigcup_{M \in \mathcal{T}_M} \left( \bigcup_{m_l \in M_l} m_l \right). \tag{2.2}$$

This grid type eases approximation of complex geometries by the unstructured tetrahedral/macro grid, and still enables high performance by exploitation of grid structure from the uniform refinement. The maximum refinement level is denoted by $L$, and mathematical quantities defined on the grid at a given refinement level are indicated by a corresponding subscript.

Specifically, we target large-scale systems with $N$, the number of DoFs, ranging up to $10^{12}$. This renders the assembly of the global matrix $A_L$ on the maximum refinement level $L$ prohibitively expensive in terms of memory. $A_L$ is therefore applied in a matrix-free manner: the *local* FE matrices $A_{m_L}$ are assembled and applied on each micro element $m_L$ on-the-fly, in a nested element-wise loop over macro and micro elements:

$$\mathbf{v}_L = A_L \mathbf{u}_L = \sum_{M \in \mathcal{T}_M} \sum_{m_L \in M_L} I_{m_L}^L A_{m_L} I_L^{m_L} \mathbf{u}_L, \quad \mathbf{u}_L, \mathbf{v}_L \in \mathbb{R}^N. \tag{2.3}$$

The loop patterns that implement these sums play a major role for performance and are described in detail in [6]. $I_L^{m_L}$ denotes the selection of DoFs located on $m_L$ from the global FE space on level $L$, while $I_{m_L}^L$ denotes the reverse: the embedding of local DoFs into global space. Those only make the notation precise, in the actual implementation, we simply load the local DoFs from a global vector without an additional operator. The local matrix $A_{m_L}$ is assembled by integrating the weak form for all combinations of $m_L$-local shape functions. This is specified in Section 3, for a general discussion of Finite Elements we refer to [14].

**2.1. Matrix-free Multigrid Components.** The large size of the linear system makes multigrid solvers an attractive choice due to their asymptotically optimal computational complexity. We point to [33] and references therein for a general introduction to multigrid methods. Here, we only highlight certain relevant components in element-wise formulations.

Multigrid methods involve mapping vectors between coarse and fine grids through an interpolation operator. We transfer a global, coarse vector $\mathbf{w}_l$ to a finer grid on level $l+1$ using a global interpolation operator $P_l^{l+1}$ with $\mathbf{w}_{l+1} = P_l^{l+1} \mathbf{w}_l$. Since we operate in a geometric multigrid setting, we are able to employ a linear interpolation. The weights in $P_l^{l+1}$ are, due to the canonical embedding of the simple linear Lagrangian FE spaces we are using, known a priori, and therefore do not need to be stored explicitly. The restriction operator is defined as the transpose of the interpolation operator.

To impement a locally-adaptive coarse-grid correction, we also require a *local* interpolation $P_{m_l}^{m_{l+1}}$, which transfers DoFs on a coarse element $m_l$ to one of its finer child elements $m_{l+1}$:

$$(2.4) \qquad \mathbf{w}_{m_{l+1}} = P_{m_l}^{m_{l+1}} I_l^{m_l} \mathbf{w}_l, \quad P_{m_l}^{m_{l+1}} \coloneqq I_{l+1}^{m_{l+1}} P_l^{l+1} I_{m_l}^l.$$

In contrast to the local problem matrix $A_{m_l}$ from (2.3), the local interpolation $P_{m_l}^{m_{l+1}}$ is *fully-assembled*, meaning it constitutes a certain selection of the global interpolation $P_l^{l+1}$ mapping between certain DoFs, not an additive component.

Furthermore, multigrid uses representations of the problem operator $A_L$ on coarser grids to obtain computationally cheap, coarse approximations to the solution. One way to build the coarse-grid operators $A_l$, $l < L$ in a geometric setting is by simply discretizing the considered operator on the coarser grids the same way as on the finest grid. This approach is commonly called direct coarse-grid approximation (DCA) or re-discretization. In a matrix-free FE environment, the local coarse-grid matrices $A_{m_l}^{\text{DCA}}$ are computed via integrals over the coarser grid elements:

$$(2.5) \qquad l < L: \quad \left[A_{m_l}^{\text{DCA}}\right]_{i,j} \coloneqq \int_{m_l} G(\mathbf{x}, \phi_i, \psi_j) d\mathbf{x},$$

where $G(\mathbf{x}, \phi_i, \psi_j)$ denotes the integrand of the weak form with shape functions from trial and test space as arguments, specified in Section 3. $i, j$ index the shape functions associated with the element $m_l$. This process involves evaluating a coefficient function potentially appearing in the weak form, on the coarser grids. DCA may fail to provide an adequate coarse-grid correction in the presence of strongly varying or discontinuous coefficients [33]. As a result, multigrid convergence can deteriorate significantly, or the method may even diverge. We demonstrate exactly when this happens in Section 4.

Two ingredients are commonly applied to recover fast convergence: Galerkin coarse-grid approximation (GCA) and operator-dependent interpolation. The Galerkin coarse-grid operator is computed as $A_{l-1}^{\text{GCA}} = (P_{l-1}^l)^T A_l P_{l-1}^l$. It turns out that for many setups with constant coefficients and canonical choices for the interpolation, DCA and GCA operators coincide. However, this is not true for the general case, especially for strongly varying coefficients. Additionally, an operator-dependent interpolation operator $P_{l-1}^l = P_{l-1}^l(A_l)$ may be required, too, to obtain satisfactory multigrid convergence rates. The design of efficient and effective interpolation operators lies at the heart of algebraic multigrid methods and is key to their efficiency [29, 30]. While extensions to operator-dependent interpolation are possible, for this paper, we will focus on Galerkin coarsening paired with linear interpolation. We comment on this in Remark 7 and refer to [33] for more details on Galerkin coarsening and operator-dependent interpolation.

**2.2. Distributed Galerkin Coarsening.** GCA, while effective for robust convergence, has disadvantages. Depending on the interpolation operator, the non-zero density of the resulting coarse grid matrices can increase rapidly [15]. Furthermore, Galerkin coarsening may be complicated to implement in parallel, can be expensive to compute (due to the triple matrix-product), cannot easily be executed matrix-free (as opposed to DCA), and requires storing the coarse grid matrices. This can be a bottleneck, especially in the massively parallel setting.

Contrary to most standard implementations using globally assembled sparse matrices, we form the Galerkin coarse-grid operator $A_l^{\text{GCA}}$ in an elementwise, distributed fashion, by decomposition of the triple-matrix product:

$$
\begin{align}
(2.6) \quad l < L: \quad A_l^{\text{GCA}} &:= (P_l^{l+1})^T A_{l+1} P_l^{l+1} \\
&= \sum_{M \in \mathcal{T}_M} \sum_{m_l \in M_l} I_{m_l}^l A_{m_l}^{\text{GCA}} I_l^{m_l} \\
(2.7) \quad \text{with} \quad A_{m_l}^{\text{GCA}} &:= \begin{cases} \sum_{m_{l+1} \in m_l} (P_{m_l}^{m_{l+1}})^T A_{m_{l+1}}^{\text{GCA}} P_{m_l}^{m_{l+1}}, & \text{if } l < L, \\ A_{m_l}, & \text{if } l = L. \end{cases}
\end{align}
$$

In (2.6), we loop over macro elements in $\mathcal{T}_M$ and the micro elements $m_l$ of the current, $l$-times refined macro element. We additively form $A_{m_l}^{\text{GCA}}$, the local Galerkin matrix on element $m_l$, in a third, innermost loop over the finer child elements $m_{l+1}$ of the current coarse element $m_l$. To that end, we interpolate locally to each fine $m_{l+1}$ with $P_{m_l}^{m_{l+1}}$ from (2.4), apply the finer-grid matrix $A_{m_{l+1}}^{\text{GCA}}$, restrict back through $(P_{m_l}^{m_{l+1}})^T$ and add up the result. Since $A_{m_l}^{\text{GCA}}$ is a local matrix, it has to be accompanied by global-local mappings $I_l^{m_l}$ and $I_{m_l}^l$ to compose the global operator $A_l^{\text{GCA}}$. The local finer-grid matrix $A_{m_{l+1}}^{\text{GCA}}$ is just the finest-grid matrix $A_{m_l}$ if $l = L - 1$, and another GCA matrix otherwise.

REMARK 1. *The spatially distributed formulation of (2.6) and (2.7) allows each process to build the coarse-grid operator independently on its respective sub-domain. This enables an embarrassingly parallel implementation of Galerkin coarsening without communication. Furthermore, it enables forming the coarse-grid operator in a locally adaptive manner.*

Unfortunately, Galerkin coarsening is difficult to incorporate efficiently in matrix-free multigrid, due to the chain of dependency to the next-finer operator. If $l = L - 1$, (2.6) can be executed matrix-free, with $A_{m_{l+1}}$ being the local matrix on the finest level. However, the resulting coarse-grid operator would be as expensive to apply as the fine-grid operator $A_L$, because $A_{m_l}^{\text{GCA}}$ is not compiled to a coarse-grid local matrix, but remains a loop with fine-grid level computational effort. By a telescope argument, if $l < L - 1$ and $A_{l+1}$ up to $A_{L-1}$ are formed with Galerkin coarsening, the coarse-grid operator at $A_l$ is as expensive to apply as the finest grid operator, defeating the efficiency of multigrid. Therefore, we compile and store the local Galerkin matrices $A_{m_l}^{\text{GCA}}$ for each micro-element for $l < L$ (i.e., for all grids but the finest grid). We discuss the memory requirements in Section 5.

**2.3. Adaptivity.** The elementwise loop (2.6) consists of localized operations, providing the flexibility to implement a spatially heterogeneous operator blending DCA and GCA. To ensure an effective coarse-grid correction, we need to avoid potentially inaccurate DCA where the coefficient exhibits strong variations, and apply Galerkin coarsening in such regions. In regions of low variation on the other hand, GCA and DCA coarse-grid operators match, and we can default to memory-efficient, matrix-free DCA operators there. For problems where strong variations are restricted to subdomains, the majority of the domain can then be treated with matrix-free DCA operators on all grid levels.

The same uniform grid refinement is applied to each macro element $M \in \mathcal{T}_M$, but we can choose and change the operator coarsening strategy depending on the macro element. Block-structured tetrahedral grids are not strictly necessary for an adaptive coarsening, but simplify its implementation due to the independent treatment of macro elements, which is already required for parallelization.

We select the macro elements that need to be treated with GCA via inspection of the coefficient of the PDE. After interpolation into a polynomial FE function, its gradient is cheaply computed. We evaluate the norm of the gradient on each micro element and apply GCA on macro elements, where the gradient inside of at least one micro element exceeds a threshold value $\nu$. We partition $\mathcal{T}_M$ into two disjoint sets of macro-elements

$$(2.8) \qquad \mathcal{T}_M^{\text{GCA}} := \{M \in \mathcal{T}_M : \max_{\substack{x \in m \\ m \in M}} \|\nabla \eta(x)\| > \nu\}, \quad \mathcal{T}_M^{\text{DCA}} = \mathcal{T}_M \setminus \mathcal{T}_M^{\text{GCA}}.$$

Finally, for the problem operator $A_l$, we formulate *adaptive* Galerkin coarse-grid approximation (AGCA):

(2.9)
$$A_l^{\text{AGCA}} = \begin{cases} \displaystyle\sum_{M \in \mathcal{T}_M^{\text{DCA}}} \sum_{m_l \in M_l} I_{m_l}^l A_{m_l}^{\text{DCA}} I_l^{m_l} + \sum_{M \in \mathcal{T}_M^{\text{GCA}}} \sum_{m_l \in M_l} I_{m_l}^l A_{m_l}^{\text{GCA}} I_l^{m_l}, & \text{for } l < L \\ \displaystyle\sum_{M \in \mathcal{T}_M} \sum_{m_l \in M_l} I_{m_l}^l A_{m_l} I_l^{m_l}, & \text{for } l = L \end{cases}$$

using the local DCA and GCA matrices from (2.5) and (2.7), respectively. $I_l^{m_l}$ denotes the purely notational mapping from the global FE space on level $l$ and the DoFs located on micro element $m_l$.

REMARK 2. *Depending on the mesh, coefficient, and choice of the threshold parameter $\nu$, (2.9) yields coarse-grid operators comparable in convergence speed to using GCA on all elements and memory consumption close to pure DCA.*

The effectiveness of AGCA is determined by the coefficient. If there are large jumps distributed over the whole domain, such that they cut every macro element, the full Galerkin coarse-grid operator (2.6) is recovered. Still, in any case, the entire fine grid is handled matrix-free. However, if the jumps are restricted to certain areas of the domain, the fraction of macro elements they cut through decreases with increasing macro-grid refinement (shown in Figure 3 and Section 5), such that the more we refine the macro grid, the fewer elements (relative to all elements) have to be Galerkin-coarsened. In fact, this paper is motivated by applications from the geosciences, particularly the simulation of Earth mantle convection, where the strongest viscosity variations occur along thin spherical shells and are therefore extremely localized [27, 4]. For such cases, we can expect only having to apply GCA on a small fraction of macro elements.

**3. Demonstrator: Stokes-Sinker Problems.** The presented adaptive coarsening scheme is largely agnostic to the underlying PDE. Problems with jumping coefficients are commonly encountered in many applications in computational science and engineering, as outlined in Section 1.2. As a concrete and challenging demonstrator for this article, we choose the variable-coefficient Stokes equations. The motivation for this choice arises from demanding geophysical applications, such as mantle convection modeling, where the Stokes equations governing momentum and mass conservation are notoriously difficult to solve due to strong radial and lateral viscosity contrasts [27, 4, 8]. This choice also demonstrates that the approach is not restricted to scalar problems with symmetric positive definite system matrices, but can also be used as a preconditioner for vector-valued problems with saddle point structure.

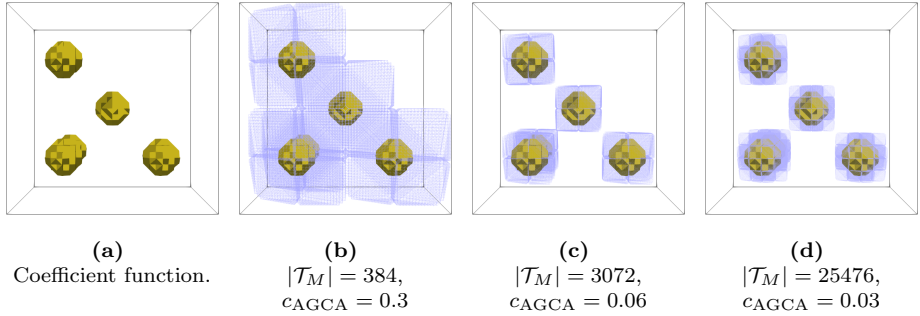

**(a)**
Coefficient function.

**(b)**
$|\mathcal{T}_M| = 384$,
$c_{\text{AGCA}} = 0.3$

**(c)**
$|\mathcal{T}_M| = 3072$,
$c_{\text{AGCA}} = 0.06$

**(d)**
$|\mathcal{T}_M| = 25476$,
$c_{\text{AGCA}} = 0.03$

**Figure 3** *Visualization of AGCA. A highly varying coefficient function is considered, with regions of high values marked in yellow. On tetrahedral elements exhibiting a steep coefficient gradient, localized GCA is applied, indicated in violet wireframe. All elements in the remaining domain are treated using matrix-free DCA. As the grid is refined, the fraction of GCA elements, $c_{AGCA}$, decreases rapidly. Consequently, multigrid robustness can be substantially enhanced with only a small number of GCA elements.*

**3.1. Variable-coefficient Stokes equations.** The variable-coefficient Stokes equations are defined as

$$
\begin{aligned}
-\nabla \cdot (2\eta(\mathbf{x})\epsilon(\mathbf{u})) + \nabla p &= \mathbf{f} \text{ in } \Omega, \\
-\nabla \cdot \mathbf{u} &= 0 \text{ in } \Omega, \\
\mathbf{u} &= \mathbf{g} \text{ on } \partial\Omega,
\end{aligned}
\tag{3.1}
$$

with velocity $\mathbf{u}$, pressure $p$, the symmetric gradient $\epsilon(\mathbf{u}) := \frac{1}{2}(\nabla\mathbf{u} + \nabla\mathbf{u}^T)$, the strongly varying or discontinuous viscosity $\eta(\mathbf{x})$, a forcing term $\mathbf{f}$, and a boundary function $\mathbf{g}$.

The weak form is derived by testing the strong form (3.1) with once weak-differentiable 0-trace functions $\mathbf{u} \in \mathbf{H}^1(\Omega) := \{\mathbf{w} \in [H^1(\Omega)]^3 : \mathbf{w} = \mathbf{0} \text{ on } \partial\Omega\}$ for the velocity and $L^2$-integrable functions $q$ for the pressure. After integration over the domain $\Omega$ and integration-by-parts, the weak formulation reads: find $\mathbf{u} \in \mathbf{H}^1(\Omega)$ and $p \in L^2(\Omega)$ with

$$
\begin{aligned}
\int_\Omega 2\eta(\mathbf{x})\epsilon(\mathbf{u}) : \epsilon(\mathbf{v}) - \int_\Omega p\nabla \cdot \mathbf{v} &= \int_\Omega \mathbf{f} \cdot \mathbf{v} &&\text{for all } \mathbf{v} \in \mathbf{H}^1(\Omega), \\
\int_\Omega q\nabla \cdot \mathbf{u} &= 0 &&\text{for all } q \in L^2(\Omega).
\end{aligned}
\tag{3.2}
$$

We use piecewise polynomial, linear velocity and pressure spaces $\mathbb{V}_l \subset \mathbf{H}^1(\Omega), \mathbb{Q}_l \subset L^2(\Omega)$ to approximate the continuous solution variables in (3.2). For the ($\mathbb{P}_1$-iso-$\mathbb{P}_2$, $\mathbb{P}_1$) pairing [5] in $d$ dimensions

$$
\mathbb{V}_l := (\mathbb{P}_1(\mathcal{T}_{m,l}))^d, \quad \mathbb{Q}_l := \mathbb{P}_1(\mathcal{T}_{m,l-1}),
\tag{3.3}
$$

the velocity is discretized on a refined grid, providing enough DoFs relative to the pressure space to guarantee inf-sup stability. An argument for the use of this discretization is given in Appendix A.

We end up with the linear system

$$(3.4) \qquad K_l \mathbf{x}_l = \begin{bmatrix} A_l & B_l^T \\ B_l & 0 \end{bmatrix} \begin{bmatrix} \mathbf{u}_l \\ \mathbf{p}_l \end{bmatrix} = \begin{bmatrix} \mathbf{f}_l \\ 0 \end{bmatrix} = \mathbf{b}_l, \quad l \in [1, \dots L]$$

$$A_l \in \mathbb{R}^{N(\mathbb{V}_l) \times N(\mathbb{V}_l)}, \quad B_l \in \mathbb{R}^{N(\mathbb{Q}_l) \times N(\mathbb{V}_l)},$$

with $N(\mathbb{S})$, the number of DoFs in the space $\mathbb{S}$. $A_l$ and $B_l$ denote the discrete viscous block and divergence, respectively, including integrals over the block-structured tetrahedral grid:

$$(3.5) \qquad [A_l]_{ij} = \int_{\mathcal{T}_{m,l}} 2\eta(\mathbf{x})\epsilon(\mathbf{\Phi}_i) : \epsilon(\mathbf{\Phi}_j), \quad \mathbf{\Phi}_i, \mathbf{\Phi}_j \in \mathbb{V}_l$$

$$[B_l]_{kj} = \int_{\mathcal{T}_{m,l}} \psi_k \nabla \cdot \mathbf{\Phi}_j, \quad \psi_k \in \mathbb{Q}_l.$$

The saddle-point system (3.4) is to be solved on $l = L$, the highest refinement level, but multigrid also operates on coarser discretizations $l < L$.

**3.2. Viscosity functions.** We aim to study the convergence properties of AGCA using a range of viscosity functions. To that end, we define the viscosity and right-hand-side in (3.1) for "sinker-like" test problems: one or multiple high-density objects sink in a medium of lower density, pushing fluid down below them that then recirculates and travels upwards in the closed domain. The contrasts in viscosity render such problems difficult to solve. The described viscosity functions and PDE solutions are visualized in Figure 4 and Figure 5, respectively.

We consider several strongly varying and discontinuous parameterized coefficient functions $\eta \in L^2(\Omega)$. Let $\mathbf{x} = (x, y, z)^T$ and $\mathrm{DR} \in \mathbb{R}$ indicate the *dynamic ratio*, a problem parameter inducing high and a low viscosity niveaus $\eta_{\mathrm{high}} = \mathrm{DR}^{0.5}$ and $\eta_{\mathrm{low}} = \mathrm{DR}^{-0.5}$. $\omega \geq 1$ denotes a smoothing parameter and $B_r^n(p)$ the set of points in distance $r$ from a point $p$, where the distance is defined by a norm $n$ ($n = 2$ for a ball, $n = \infty$ for a cube). We aim for some coefficient shapes to be unaligned with a refined unit cube mesh on all grid levels, which we center around $C_{\mathrm{ua}} = \frac{11}{24}$, to guarantee grid-unalignment. With that, we define characteristic functions $\xi_i$:

$$\xi_1(\mathbf{x}) := \mathbf{1}_{B_{\frac{1}{8}}^\infty((\frac{1}{2}, \frac{1}{2}, \frac{1}{2}))} \qquad \xi_2(\mathbf{x}) := \mathbf{1}_{B_{\frac{1}{8}}^\infty((C_{\mathrm{ua}}, C_{\mathrm{ua}}, C_{\mathrm{ua}}))}$$

$$\xi_3(\mathbf{x}) := \xi_{4\mathrm{a}}(x, C_{\mathrm{ua}}, \tfrac{1}{8}) \cdot \xi_{4\mathrm{a}}(y, C_{\mathrm{ua}}, \tfrac{1}{8}) \cdot \xi_{4\mathrm{a}}(z, C_{\mathrm{ua}}, \tfrac{1}{8}) \qquad \xi_4(\mathbf{x}) := \mathbf{1}_{B_{0.1}^2((\frac{1}{2}, \frac{1}{2}, \frac{1}{2}))}$$

$$\xi_5(\mathbf{x}) := \prod_{j=1}^{n_{\mathrm{sinkers}}} \xi_{5\mathrm{a}}(\mathbf{x}, \mathbf{p}_j, 0.1) \qquad \xi_6(\mathbf{x}) := \prod_{j=1}^{n_{\mathrm{sinkers}}} \mathbf{1}_{B_{0.1}^2(\mathbf{p}_j)}$$

(3.6)

using the auxiliary functions

$$(3.7) \qquad \xi_{3\mathrm{a}}(x, p, r) := \frac{1}{2}[\tanh(\omega \cdot (x - (p - r))) - \tanh(\omega \cdot (x - (p + r)))]$$

$$(3.8) \qquad \xi_{5\mathrm{a}}(\mathbf{x}, \mathbf{p}, r) := 1 - \exp(-\omega \cdot \max(0, ||\mathbf{p} - \mathbf{x}||_2 - \frac{r}{2})^2).$$

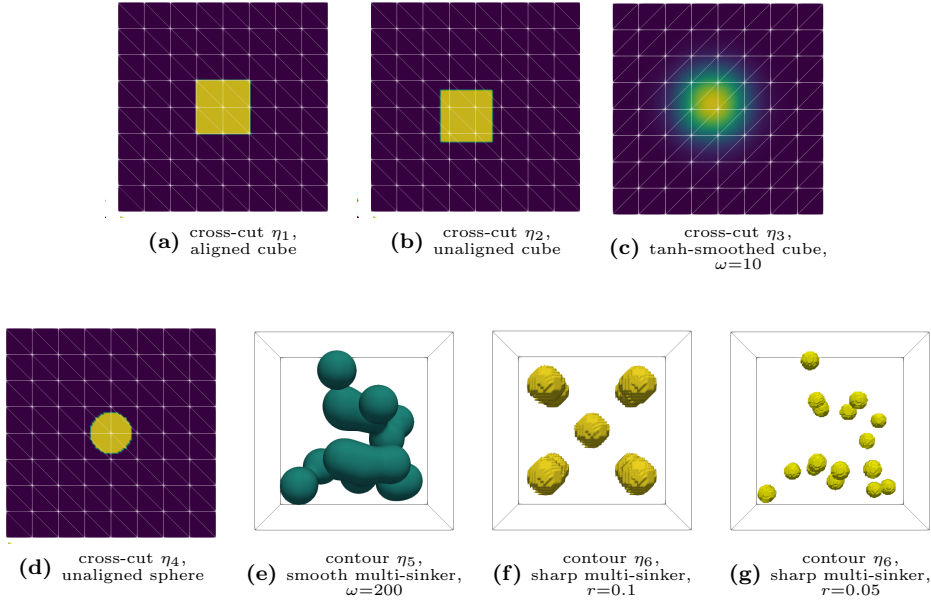

(a) cross-cut $\eta_1$, aligned cube
(b) cross-cut $\eta_2$, unaligned cube
(c) cross-cut $\eta_3$, tanh-smoothed cube, $\omega=10$
(d) cross-cut $\eta_4$, unaligned sphere
(e) contour $\eta_5$, smooth multi-sinker, $\omega=200$
(f) contour $\eta_6$, sharp multi-sinker, $r=0.1$
(g) contour $\eta_6$, sharp multi-sinker, $r=0.05$

*Figure 4* **Visualizations of the test coefficients in (3.6)- (3.9).** *Yellow marks areas where $\eta = \eta_{high}$, white and violet marks areas where $\eta = \eta_{low}$. The functions $\eta_1 - \eta_4$ are visualized by cross cuts through the unit cube, functions $\eta_5, \eta_6$ as 3D contour plots. We consider grid-aligned and unaligned cuboid ($\eta_1, \eta_2$), as well as unaligned spherical discontinuities ($\eta_4, \eta_6$) and smooth coefficients with large gradients ($\eta_3, \eta_5$). The multi-sinker problem can be implemented either with an exponential decay at the sinker boundaries, indicated here by a decreasing green blurring, or with a sharp jump.*

Then, the viscosity functions $\eta_i$ are obtained by rescaling the characteristic functions:

$$\eta_i(\mathbf{x}) := \eta_{\text{low}} + (\eta_{\text{high}} - \eta_{\text{low}}) \cdot \xi_i \quad \text{for } i \in \{1,2,3,4,6\}, \tag{3.9}$$

$$\eta_5(\mathbf{x}) := \eta_{\text{low}} + (\eta_{\text{high}} - \eta_{\text{low}}) \cdot (1 - \xi_5) \tag{3.10}$$

with corresponding right-hand-sides in the Stokes equation defined as:

$$\mathbf{f}_i := (0, 0, -\xi_i) \quad \text{for } \in \{1,2,3,4,6\} \tag{3.11}$$

$$\mathbf{f}_5 := (0, 0, -(\xi_5 - 1)) \tag{3.12}$$

The right-hand-side implements a downward drag at the location of high viscosity. We combine this to an enclosed flow with homogeneous Dirichlet boundary conditions ($\mathbf{g} = 0$).

A common idea to improve the convergence of a linear solver in the presence of discontinuous or sharply varying coefficients involves approximations of the coefficient [11, 32]. By interpolating the coefficient into the FE-space, we essentially move from a sharp, possibly unaligned jump to, e.g., a piecewise-linear, continuous function with large gradient, or a piecewise constant, discontinuous function. During local assembly, we then cheaply evaluate a polynomial at the quadrature points. For the experiments below, additionally to standard analytical evaluation, we consider four types of coefficient approximations: interpolation into the space of continuous, piecewise linear functions $\mathbb{P}_1$, and three types of projections (arithmetic, harmonic, and geometric mean) into the space of piecewise constant, possibly discontinuous functions $\mathbb{P}_0$. Figure 6 illustrates the effect of these approximations for a one-dimensional example on three elements.

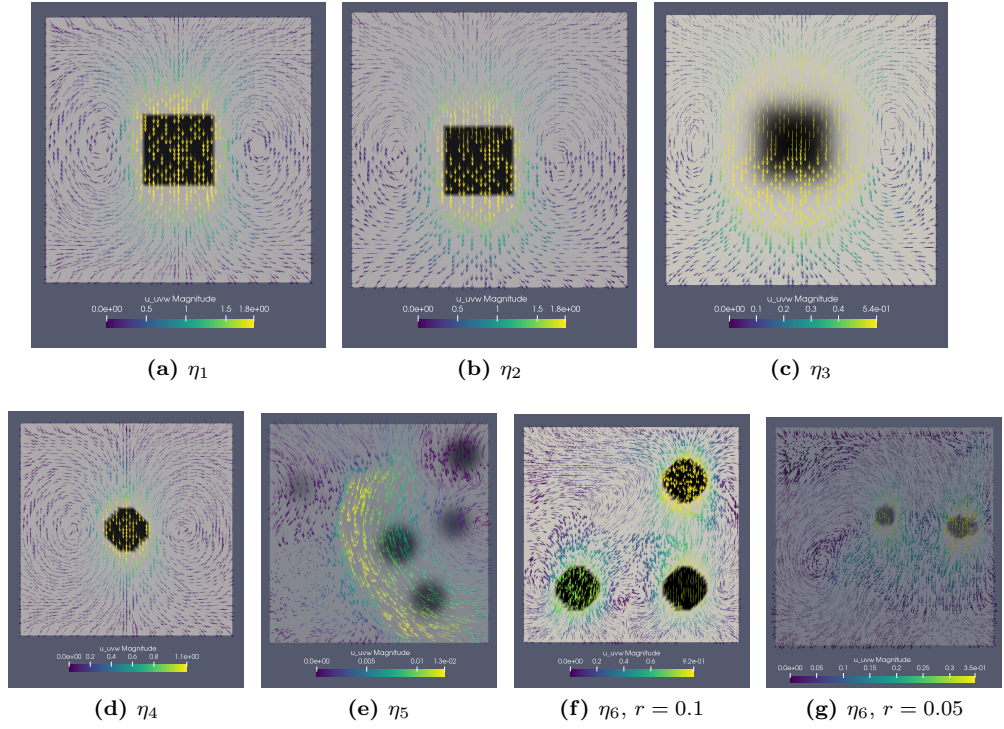

**Figure 5** *2D Crosscuts of the velocity solutions for the problems defined by Stokes equation (3.1) with viscosities (3.9) and right-hand-sides (3.11).*

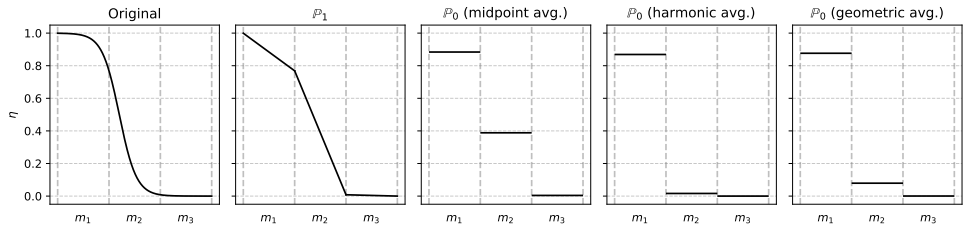

**Figure 6** *Illustration of the different averaging techniques applied to a smooth but steep 1D coefficient over three elements $m_1, m_2, m_3$.*

REMARK 3. *We use second order quadrature rules to evaluate the integrals of the weak formulation. However, for optimal error convergence of the discretization, 0-th degree quadrature may be sufficient due to the linear polynomial spaces.*

**4. Algebraic Convergence.** We incorporate the AGCA-coarsening in a Multigrid preconditioner for a flexible generalized minimal residual method (FGMRES) [14] outer solver. This is necessary due to the use of a non-symmetric triangular preconditioner for the linear system (3.4):

$$(4.1) \qquad Q_L^{-1} K_L x_L = Q_L^{-1} b_L, \quad Q_L^{-1} := \begin{bmatrix} \tilde{A}_L^{-1} & \tilde{A}_L^{-1} B_L^T \tilde{S}_L^{-1} \\ 0 & \tilde{S}_L^{-1} \end{bmatrix}.$$

$\tilde{A}_L^{-1}$ denotes the approximate inverse of the velocity block, implemented as a single AGCA-multigrid V-cycle. For the smoother, we apply a Chebychev polynomial smoother of order 2-4 and up to two pre/post smoothing steps. The coarsest-grid systems are solved by a conjugate gradient method (CG) preconditioned by successive over-relaxation (SOR). The high variations in the viscosity make the use of an advanced Schur complement preconditioner necessary [28]. We use the diag($A$)-BFBT least-squares commutator [13]:

$$(4.2)\\ \tilde{S}_L^{-1} := Z^{-1} \cdot (B_L W^{-1} A_L W^{-1} B_L^T) \cdot Z^{-1}, \quad W := \mathrm{diag}(A_L), \quad Z := B_L W^{-1} B_L^T.$$

Additional information on the solver design and implementation is given in Appendix B. We compute on a comparatively small problem size of $6.7 \times 10^6$ DoFs from 48 macro-elements and refinement level 5, leaving scaling considerations to Section 6.

We declare convergence once

$$(4.3) \qquad ||\mathbf{r}^{(i)}||_2 \leq 10^{-6} \cdot ||\mathbf{b}||_2,$$

or the residual reaches double-accuracy machine precision.

REMARK 4. *This criterium is widely used in the literature and provides basic comparability of different solvers. However, in our case, the right-hand-sides are already rather small, $||\mathbf{b}||_2$ is $O(10^{-8})$, such that this criterion forces solving until the residual is close to float64 machine precision. With a decrease of the residual by just three magnitudes, the residual is $O(10^{-11})$, the solutions stay visually unchanged as depicted in Figure 5.*

**4.1. Direct Coarse-grid Approximation.** Figure 7 demonstrates, under which circumstances DCA coarse-grid operators (2.5) are not effective and outer solver convergence suffers. This also includes the viscosity evaluation techniques presented in Section 3, namely interpolation to an FE function and mean, harmonic, and geometric averaging.

Figure 7, $\eta_1$: Using appropriate quadrature rules (for linear elements, that would be rules that do not evaluate on the element boundary), a jump that is aligned with the elements of the macro subdivision can always be represented exactly on the coarser grids. The jump does not cut macro nor micro elements, such that the coefficient is constant locally on each element. Thereby, the coefficient evaluation is independent of the grid resolution, such that DCA provides an accurate representation of the fine-grid operator on the coarser grids. All variants converge fast with only light dependence on the dynamic ratio.

Figure 7, $\eta_2$ and $\eta_4$: When translating the cube to an unaligned position, or using a spherical shape, the jump plane becomes unaligned with the grid. Such viscosities feature discontinuities within elements. In contrast to a grid-aligned jump, the refinement level on which the jump is evaluated makes a major difference for the resulting FE matrix: the strength of connection of DoFs lying close to the jump edge in the global matrix vastly changes between refinement levels. Thereby, DCA evaluates

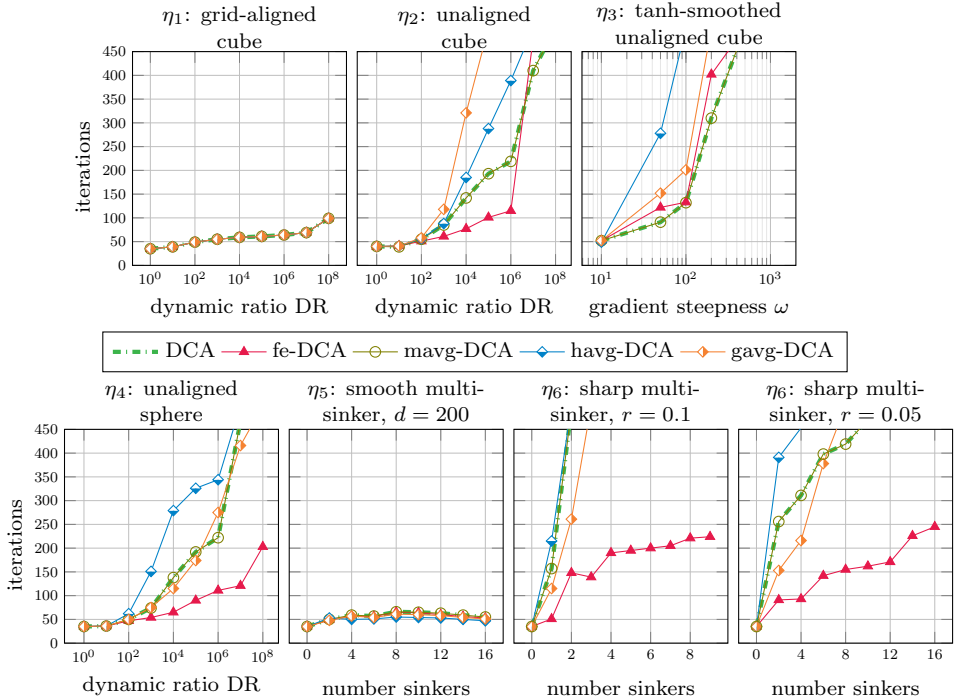

**Figure 7** *Multigrid-preconditioning using direct coarse-grid approximation (DCA). Convergence results in number of FGMRES iterations against various problem parameters like dynamic ratio, gradient steepness and number of sinkers. The sharp multi-sinker experiments are conducted with a dynamic ratio of $10^6$.*

the coefficient inaccurately on the coarser grids, the coarse-grid correction does not match the finest-grid problem anymore, and the resulting multigrid preconditioners become ineffective with higher dynamic ratio. Interpolating the coefficient into an FE-function shows the least deterioration, as it replaces the jump plane with a smooth function, and still yields acceptable iteration numbers up to variations of $10^6$.

Figure 7, $\eta_3$: We smooth out the unaligned, cuboid jump from $\eta_2$ with a tanh function, yielding $\eta_3$. With more coefficient smoothing and low gradient steepness $\omega$, the tanh can be resolved accurately on all grids with low variation per element, and convergence is fast. With a higher gradient steepness, the analytically smooth function shows high variation also within a single element and between quadrature points. In the discrete sense, it appears as a jumping function. Consequently, convergence suffers analogously to the unaligned cube ($\eta_2$).

Figure 7, $\eta_5$: We increase the area of the domain that experiences a large variation in the coefficient by adding several unaligned, but smooth sinkers. A smooth multi-sinker problem with a viscosity contrast of $10^6$ but only a light exponential decay $\delta = 200$ can be resolved similarly well on all grids. Although we encounter a high variation, the change of viscosity within an element is small, similar to the lower end of gradient steepness in $\eta_3$. We observe fast convergence independent of the number of sinkers.

Figure 7, $\eta_6$: By changing the smooth decay at the sinker boundary to a sharp jump, the problem difficulty increases drastically. DCA approaches slow down consid-

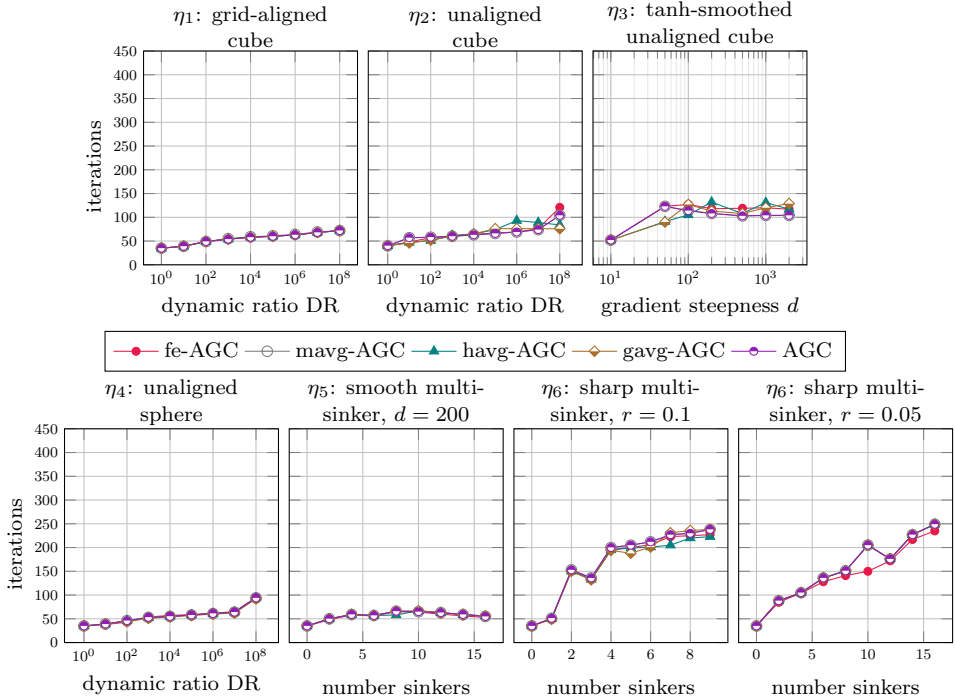

*Figure 8   Multigrid-preconditioning using adaptive Galerkin coarse-grid operators (AGCA). Convergence results in number of FGMRES iterations against various problem parameters like dynamic ratio, gradient steepness and number of sinkers. We fix the AGCA gradient tolerance $\nu$ to 10 for all experiments. The sharp multi-sinker experiments are conducted with a dynamic ratio of $10^6$.*

erably already with a single sinker and practically stop converging with an increasing number of sinkers. Smaller sinkers can be placed in higher number in the domain without convoluting it. The resulting problem seems to be easier to solve, as convergence breaks down later. Interpolating the coefficient into an FE-function shows the least deterioration, as it replaces the jump plane with a smooth function, and still yields acceptable iteration numbers up to variations of $10^6$.

**4.2. Adaptive Galerkin Coarse-grid Operators.** Figure 8 shows, that AGCA applied to the discrete Stokes equation leads to a robust multigrid preconditioner, also for cases where pure DCA fails. We fix the AGCA gradient tolerance $\nu$ in (2.9) to 10.

Over all test cases, we observe that the choice of coefficient approximation does not have a large influence on the convergence curve when using adaptive Galerkin coarsening. This is caused by evaluations on multiple grids with DCA yielding vastly different results with different coefficient approximations. AGCA on the other hand relies on a single, finest-grid coefficient evaluation, that is passed down to coarser grids.

Figure 8, $\eta_1$: For the grid-aligned jump, AGCA and DCA coarse-grid operators are equivalent, as the coefficient is also aligned with all coarse grids and thereby takes exactly the same shape on all grids. For such problems, DCA is sufficient, and AGCA operators simplify to pure DCA operators.

Figure 8, $\eta_2, \eta_3, \eta_4$: Galerkin coarse-grid operators implicitly map back to the finest grid and apply the finest-grid operator. Thereby, they also evaluate the coefficient on the finest grid and produce a coarse-grid correction matching the finest-grid problem. Contrary to DCA, the resulting multigrid solvers converge with only a light dependence on the dynamic ratio also for the unaligned cube and sphere, *without the need to smooth out the coefficient.*

Figure 8, $\eta_5$: On the smooth multi-sinker problems, DCA and AGCA are very similar in convergence speed. The coefficient is smooth enough such that its representation on coarser grids matches well with that on the finest grid. Depending on the tolerance $\nu$ in (2.9), Galerkin coarse-grid operators are only formed on few elements of highest variation in the coefficient. On most other elements, AGCA simplifies to DCA.

Figure 8, $\eta_6$: In the sharp multi-sinker problems, DCA solvers essentially stop to converge with only few sharp sinkers, if the coefficient is not smoothed by interpolating into a FE-function. Although the iteration count increases moderately with an increasing number of sinkers, AGCA maintains robust convergence for all coefficient approximations.

**4.3. Adaptive Coarsening Tolerance $\nu$.** It depends on the coarsening tolerance $\nu$ in (2.9) whether a macro element is Galerkin-coarsed. Choosing a higher tolerance allows elements with high gradient in the viscosity be treated by DCA, which can destroy convergence. Choosing a very small tolerance ensures fast convergence but may lead to a higher memory demand due to more stored Galerkin coarse-grid matrices. However, even for a low tolerance, constant coefficient/0 gradient areas are still treated by DCA. Figure 9 demonstrates how AGCA blends DCA and GCA depending on the coarsening tolerance $\nu$.

**5. Memory.** This section aims to quantify the potential memory savings from using matrix-free operators and adaptive coarsening against a matrix-based approach.

A global vector on the finest grid holding $N_L := N(\mathbb{V}_L) + N(\mathbb{Q}_L)$ DoFs serves as the basic unit of measurement here. We convert velocity and pressure DoFs to global DoFs $N_L$ via tetrahedral numbers from [23]:

$$c_u := 1 - \lim_{l \to \infty} \frac{N_{\text{vertices}}(l)}{3 \cdot N_{\text{vertices}}(l+1)} \approx 0.95, \quad N(\mathbb{V}_L) = c_u \cdot N_L$$
(5.1)
$$c_p := 1 - c_u \approx 0.05, \quad N(\mathbb{Q}_L) = c_p \cdot N_L.$$

To solve the discrete Stokes problem (3.4), we require two global vectors $x_L, b_L$ for solution and right-hand-side, each holding $N_L$ DoFs. Within the diag($A$)-BFBT (4.2), the CSR matrix for the suboperator $Z$ takes up $\sim 4 \cdot N_L$. Furthermore, The FGMRES outer solver requires $(1 + n_{\text{restart}}) \cdot 2 \cdot N_L$ memory.

For reference, we assume a global sparse matrix assembly of the Stokes operator $K_L$ on the finest grid. We have 15 couplings connecting each velocity DoFs to DoFs on neighboring vertices in 3 cross-couplings of the velocity block $A_L$, resulting in 45 nonzeros per row for the velocity block. The gradient $B^T$ connects pressure DoFs to the 15 surrounding velocity DoFs. Together with a factor to account for 64-bit indices in the assumed CSR sparse matrix storage, row-pointers and (5.1), we obtain for the

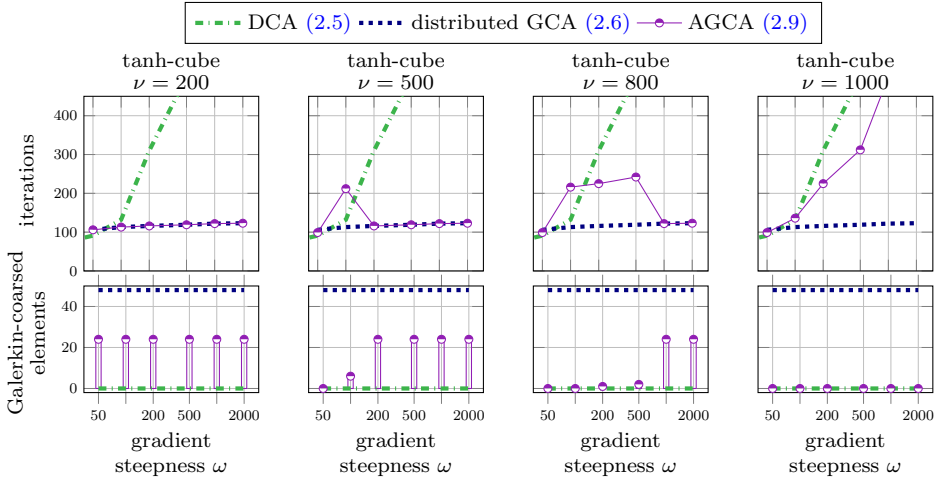

*Figure 9 Convergence speed and memory requirement over coarsening tolerance $\nu$.* We observe convergence speed and memory requirement when varing the coarsening tolerance $\nu$ in (2.9) for the tanh-smoothed cube ($\eta_3$, Figure 4c): with a small $\nu$, the condition in (2.9) triggers even with small gradient steepness $\omega$. This leads to all 24 elements in the area of the tanh-smoothed cube ($\eta_4$, Figure 4d) being Galerkin coarsed (first column). The viscosity is constant on the remaining 24 elements in this 48 element grid, such that they are treated with DCA regardless of $\nu > 0$. With low enough $\nu$, AGCA is equivalent to distributed GCA in algebraic convergence. With increasing tolerance (2.-4. column), higher gradients are tolerated and treated with DCA, leading to lower memory demand due to fewer Galerkin-coarsed elements, but deteriorating convergence. For this small setup, the memory savings from AGCA are limited to 24 macro elements due to the very small size of the coarse-grid (48 macro elements). As shown later in Figure 10, there is much larger potential savings with more macro elements.

memory load of a globally assembled $A_L$ and $K_L$:

$$
\text{Mem}_{A_L} = (\ \underbrace{3 \cdot 15}_{\substack{\text{nonzeros} \\ \text{per row} \\ \text{in } A_L}} \cdot \ \underbrace{2+1}_{\text{CSR indices}} \ ) \cdot N(\mathbb{V}_L) \approx 86.5 \cdot N_L \tag{5.2}
$$

$$
\text{Mem}_{K_L} = \text{Mem}_{A_L} + 2 \cdot (\ \underbrace{15}_{\substack{\text{nonzeros} \\ \text{per row} \\ \text{in } B_L^T}} \cdot \ \underbrace{2+1}_{\text{CSR indices}} \ ) \cdot N(\mathbb{Q}_L) \approx 89.5 \cdot N_L. \tag{5.3}
$$

For memory load of the coarse-grid operators, we assume that we only Galerkin-coarsen the velocity block $A_L$, as only there the coefficient enters. Assembling a global sparse matrix for the Galerkin coarse-grid operator involves explicitly forming the product $A_{L-1}^{\text{GCA}} = P^T A_L P$. Multiplication with the interpolation/restriction operator from both sides can lead to an additional fill-in, increasing the stencil size, depending on the type of interpolation and grid used. On a structured, tetrahedral grid and with basic linear 1-distance interpolation, we observe $n_{\text{fill-in}} \approx 1$. More involved long-range interpolations can lead to larger $n_{\text{fill-in}}$ [15]. Then, GCA sparse matrix takes up

$$
\text{Mem}_{A_{L-1}^{\text{Galerkin}}} = \underbrace{\frac{n_{\text{fill-in}} \cdot \text{Mem}_{A_L}}{8}}_{\text{fill-in, coarse-grid}} \approx n_{\text{fill-in}} \cdot 10.8 \cdot N_L. \tag{5.4}
$$

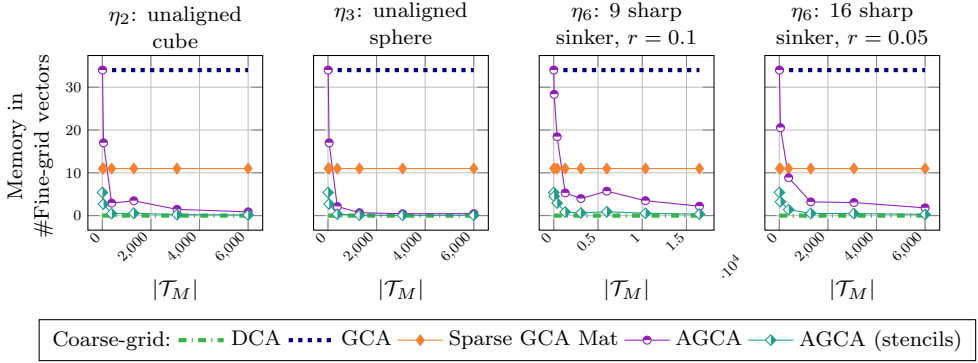

*Figure 10  Memory consumption of the coarse-grid operator over the size of the macro-grid.*
*Smooth viscosities are omitted, as for them, GCA is not required for fast convergence. The adaptive coarsening tolerance $\nu$ is set to 10 in all cases.*

The elementwise, distributed GCA formulation (2.6) stores a $4 \times 4$ matrix per element. To convert the number of elements into number of vertices, we use again tetrahedral numbers [23]:

$$\lim_{l \to \infty} \frac{N_{\text{vertices}}(l)}{N_{\text{cells}}(l)} = \frac{1}{6}, \tag{5.5}$$

implying that in the limit of infinite refinement, there are $6 \times$ more elements than vertices. We obtain for the memory load of GCA on all elements:

$$\underbrace{3}_{\substack{\text{velocity} \\ \text{components}}} \cdot \underbrace{16}_{\substack{\text{local matrix} \\ \text{entries}}} \cdot \underbrace{6}_{\text{cells to vertices}} \cdot \underbrace{\frac{1}{8}}_{\text{coarse-grid}} \cdot N(\mathbb{V}_L) \approx 34 \cdot N_L. \tag{5.6}$$

Thereby, distributed GCA on all elements would not be justified in comparison to assembling a sparse GCA matrix, due to its memory consumption. However, it enables the locally adaptive approach central to this paper. The memory load of AGCA depends on the fraction of macro elements being Galerkin-coarsed:

$$c_{\text{AGCA}} := \frac{|\{M \in \mathcal{T}_M : \ \nabla \eta(\mathbf{x})|_M \geq \nu\}|}{|\mathcal{T}_M|}, \quad c_{\text{AGCA}} \in [0, 1]. \tag{5.7}$$

Figure 10 illustrates the memory load of the coarse-grid operator as a function of the grid size. As the grid becomes larger, a smaller fraction of elements experience strong variations or jumps in viscosity. Consequently, $c_{\text{AGCA}}$ decreases with increasing problem size, and the memory load of the AGCA coarse-grid operator is reduced accordingly. In the observed cases, the memory required for the coarse-grid operators becomes negligible, while robust convergence rates are maintained. This reduction occurs more rapidly for test cases involving a single sphere or cube, but even for multi-sinker configurations, the memory consumption tends toward negligible levels.

REMARK 5. *A more efficient way to implement AGCA would be to store the coarse operators as stencils per DoF, not requiring the storage of any indexes and* *consuming (assuming $n_{\text{fill-in}} = 1$):*

$$(5.8) \qquad \underbrace{3}_{\substack{velocity \\ components}} \cdot \underbrace{15}_{\substack{stencil \\ entries}} \cdot \underbrace{\frac{1}{8}}_{coarse\text{-}grid} \cdot N(\mathbb{V}_L) \approx 5.4 \cdot N_L.$$

*This improves the worst-case of high coefficient gradients entering every macro element ($c_{AGC} \approx 1$). On the other hand, for $c_{AGC} \ll 1$, the storage method of the local coarse-grid operators is not a relevant factor anymore.*

Table 1 gives a summary comparing a fully matrix-based approach with computing matrix-free on the finest grid and assembling global GCA matrices or using AGCA on the coarser grids.

|  | purely Matrix-based | Matrix-free finest grid | | | |
|---|---|---|---|---|---|
|  |  | Sparse GCA Mat | AGCA | AGCA (stencils) | DCA |
| PDE | 2 | | | | |
| FGMRES | $2 + 2 \cdot n_{\text{restart}}$ | | | | |
| Preconditioner | 4 | | | | |
| Fine-grid | 89.5 | 0 | 0 | 0 | 0 |
| Coarse-grid | 10.8 | 10.8 | $c_{\text{AGC}} \cdot 34$ | $c_{\text{AGC}} \cdot 5.4$ | 0 |

**Table 1** *Memory consumption measured in fine-grid vectors with $N_L$ DoFs: globally assembled matrices on all grids (1. column), matrix-free on the finest grid with assembled matrices on coarser grids (2. column), adaptive Galerkin coarsening (3. column) and DCA on the coarse grids (4. column).*

REMARK 6. *Using matrix-free operators saves $89.5$ fine-grid vectors, and applying AGCA saves another $10.8$ fine-grid vectors for the observed test cases. However, the choice of outer solver is crucial for memory consumption: FGMRES may, depending on the restart value, require the largest fraction of the overall allocated memory to store the Krylov basis vectors. Then, matrix-free methods and AGCA can be seen as a way to buy a larger FGMRES restart in large-scale problems. Other Krylov variants like minimal residual method (MinRes) have a much lower, limited memory demand, but the required symmetric, block-diagonal preconditioner is observed to be less robust in our experience, leading to more outer iterations and higher time-to-solution.*

**6. Scalability.** In this section we investigate how well our solver scales. This includes its convergence rate with increasing refinement level and a weak scaling. We are conducting our scaling runs on SuperMUC-NG Phase 1, which contains Intel Skylake Xeon Platinum 8174 processors and 48 cores per node. Each node has 96 GB of main memory available.

Figure 11 shows the iteration counts of FGMRES preconditioned with AGCA multigrid in combination with diag($A$)-BFBT across different grid refinements. For the simpler cases $\eta_1$, $\eta_2$, and $\eta_4$, we observe only mild increases in iteration counts. The smooth multi-sinker problem becomes progressively easier with refinement, since the coefficient variation is distributed across more, smaller elements. Even the more challenging sharp multi-sinker case with 8 sinkers, $\eta_6$, can be solved effectively at higher refinement levels with only a modest increase in iterations. When moving to the most difficult case with 16 sharp sinkers, the iteration growth becomes more pronounced, yet the solver remains stable and convergent. For such demanding problems, a switch from linear to operator-dependent interpolation may be necessary. This can also be incorporated adaptively, as discussed in Remark 7.

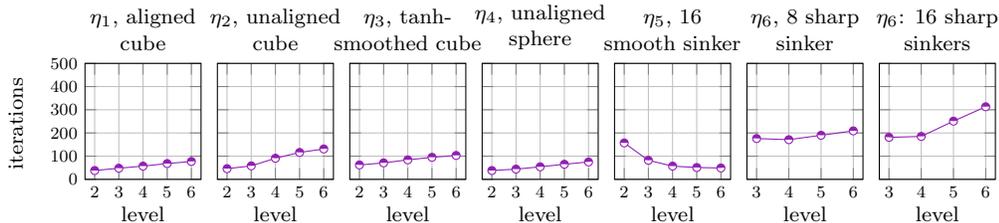

*Figure 11 FGMRES Iteration numbers over refinement level with adaptive Galerkin-coarsening as preconditioner.*

Figure 12 presents weak scaling results for the smooth multi-sinker and sharp sphere test cases, respectively. The solver is stopped at a residual reduction of $10^{-3}$; further reduction leaves the solution visually unchanged, indicating that sufficient accuracy has been reached.

In both test cases, the multigrid cycle, excluding the coarsest-grid solver, scales almost perfectly from 1,296 processes (36 nodes) with $1.8 \cdot 10^8$ DoFs up to 105,456 processes (2,930 nodes) with $1.44 \cdot 10^{10}$ DoFs. At larger problem sizes, both the $\text{diag}(A)$-BFBT preconditioner and the coarsest-grid solver exhibit a slight slowdown. The coarsest grid is solved using a SOR-preconditioned CG, which, similar to the outer solver, shows a grid-dependency in iterations. Within the $\text{diag}(A)$-BFBT, the cost of the Hypre-AMG cycle in the $Z$-solves increases modestly. Although not all solver components scale perfectly, the final time-to-solution is acceptable for a Stokes problem with $10^{10}$ DoFs and an unaligned viscosity jump of magnitude $10^6$.

Using the same solver design but employing Hypre-AMG on the velocity block instead of AGCA leads to out-of-memory errors, due to the assembly of the diffusion operators on the finest and coarser grids. In contrast, a solver utilizing DCA within the multigrid hierarchy fails to converge in a reasonable time-to-solution, as indicated in Section 4. By applying Galerkin coarse approximations selectively in the appropriate subdomains using AGCA, convergence is recovered and solver robustness is maintained, with only minimal impact on memory consumption.

REMARK 7. *Achieving truly grid-independent convergence with lower iteration counts may require a more sophisticated weighting $W$ in the Schur complement approximation (4.2), along with an upgrade from the simple linear interpolation to operator-dependent interpolation in the velocity block multigrid $\tilde{A}_L^{-1}$. The latter can also be realized in a localized, distributed way by replacing the interpolation matrices in (2.4). Furthermore, an adaptive scheme can also be employed with respect to the interpolation operator: a linear interpolation is sufficient in regions of low variation, where the weights are known and need not be stored. Only in regions of high variation, the interpolation weights are deduced from the operator and stored.*

**7. Conclusion.** We have developed a multigrid coarse-grid correction that applies Galerkin coarsening selectively in regions with strong viscosity variation, and resorts to light-weight, low-memory direct coarse-grid approximation otherwise.

The approach has achieved robust algebraic convergence even for highly heterogeneous and discontinuous viscosities, with jump magnitudes up to $10^8$ and up to 16 smooth and sharp sinkers, where pure direct coarse-grid approximation fails to provide an accurate coarse-grid correction. The convergence speed of the scheme is mostly independent of whether the coefficients are obtained through direct evaluation, interpolation to a linear finite element function, or averaging.

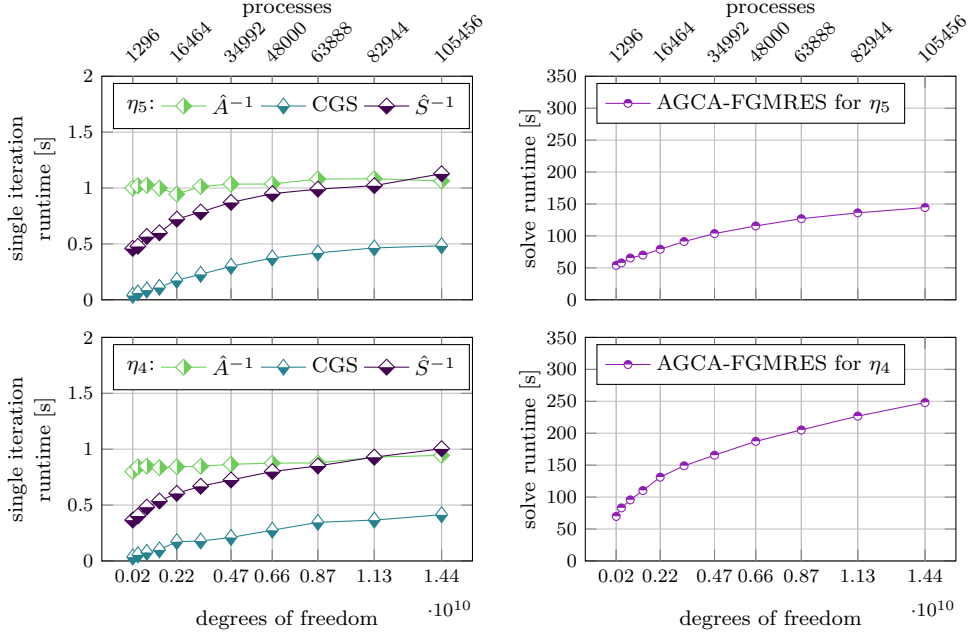

**Figure 12** *First row: Weak scaling results for the smooth multi-sinker benchmark $\eta_5$ with 16 sinkers, $DR = 10^6$, and $\omega = 200$. Second row: Weak scaling results for the single sphere $\eta_4$ with unaligned jump, $DR = 10^6$. The left column reports the timings per FGMRES iteration, broken down into the main solver components: the velocity-block multigrid $\hat{A}^{-1}$ (excluding the coarsest-grid solver), the coarsest-grid solver (CGS), and the application of the Schur complement approximation $\hat{S}^{-1}$. The right column plots present the overall time-to-solution required to reduce the residual by a factor of $10^{-3}$.*

We have quantified the potential memory savings of this scheme and demonstrated that, owing to the highly localized application of GCA, the memory cost of the coarser operators is minimized in the cases considered. However, the choice of the outer solver has proved crucial: depending on the restart parameter, FGMRES may dominate the overall memory consumption.

Finally, we have demonstrated the efficiency of the solver by solving a Stokes problem with $10^{10}$ DoFs, including an unaligned viscosity jump of magnitude $10^6$, in 250 seconds. To further improve algebraic convergence speed and eliminate a grid-dependency of convergence, switching from simple linear to operator-dependent interpolation is considered in the future.

**Acknowledgments.** The authors gratefully acknowledge funding through the joint BMBF project CoMPS[1] (grant `16ME0647K`). The authors would like to thank the NHR-Verein e.V.[2] for supporting this work/project within the NHR Graduate School of National High Performance Computing (NHR). The authors gratefully acknowledge the scientific support and HPC resources provided by the Erlangen Na-tional High Performance Computing Center (NHR@FAU) of the Friedrich-Alexander-Universität Erlangen-Nürnberg (FAU). NHR funding is provided by federal and Bavarian state authorities. NHR@FAU hardware is partially funded by the German Research Foundation (DFG) – 440719683. The authors gratefully acknowledge the Gauss Centre for Supercomputing e.V.[3] for funding this project. The authors gratefully acknowledge the Gauss Centre for Supercomputing e.V. (www.gauss-centre.eu) for funding this project by providing computing time on the GCS Supercomputer SuperMUC-NG at Leibniz Supercomputing Centre (www.lrz.de). The research was co-funded by the financial support of the European Union under the REFRESH—Research Excellence For Region Sustainability and High-tech Industries—Project No. CZ.10.03.01/00/22_003/0000048 via the Operational Programme Just Transition.

---

[1] https://gauss-allianz.de/en/project/title/CoMPS
[2] https://www.nhr-verein.de

**Appendix A. ($\mathbb{P}_1$-iso-$\mathbb{P}_2$, $\mathbb{P}_1$) Finite Element Discretization.**

The argument for this discretization compared to the common choice of lowest-order Taylor-Hood elements $\mathbb{V}_l, \mathbb{Q}_l \coloneqq (\mathbb{P}_2(\mathcal{T}_{m,l}))^d, \mathbb{P}_1(\mathcal{T}_{m,l-1})$, is the following: a jump in the coefficient most of the time implies a non-smooth solution of the continuous PDE problem. That is, if the right-hand-side does not exactly mirror the jump, the grid-convergence of the discretization of higher-order polynomial spaces degrades. The increased polynomial degree may still yield a lower constant in the discretization error, but the convergence order is the same as for a lower-order discretization like (3.3). In that case, (3.3) produces the same problem size in number of DoFs, but the resulting problem operator may be computationally cheaper to apply: due to the linear polynomials, the matrix is sparser, and local integration requires lower degree quadrature rules. This is not generally true and there are counterexamples with certain hierarchical basis functions, but holds in our case.

**Appendix B. Solver Implementation.** Since the discrete gradient and divergence operators $B_l$ and $B_l^T$ in (3.4) are not affected by the coefficient $\eta(\mathbf{x})$ we treat them with DCA on coarser grids and only apply AGCA to the viscous $A$-block. With this, the complete coarse-grid Stokes operator reads:

$$(\text{B.1}) \qquad l \leq L: \quad K_l = \begin{bmatrix} A_l^{\text{AGCA}} & (B_l^{\text{DCA}})^T \\ B_l^{\text{DCA}} & 0 \end{bmatrix}.$$

The symmetric gradient in (3.5) induces a cross-coupling of the spatial velocity components. We choose block-diagonal, linear interpolation operators $P_{m_l}^{m_{l+1}}$, where each diagonal block $P$ transfers a single velocity component between grids from a micro element $m_l$ to $m_{l+1}$. In 2D:

$$\begin{aligned}
(\text{B.2}) \quad A_{m_l}^{\text{GCA}} &= \sum_{m_{l+1} \in m_l} (P_{m_l}^{m_{l+1}})^T A_{m_{l+1}} P_{m_l}^{m_{l+1}} \\
&= \sum_{m_{l+1} \in m_l} \begin{bmatrix} P^T & 0 \\ 0 & P^T \end{bmatrix}_{m_l}^{m_{l+1}} \begin{bmatrix} A_{xx} & A_{xy} \\ A_{yx} & A_{yy} \end{bmatrix}_{m_{l+1}} \begin{bmatrix} P & 0 \\ 0 & P \end{bmatrix}_{m_l}^{m_{l+1}} \\
&= \sum_{m_{l+1} \in m_l} \begin{bmatrix} P^T A_{xx} P & P^T A_{xy} P \\ P^T A_{yx} P & P^T A_{yy} P \end{bmatrix}_{m_{l+1}},
\end{aligned}$$

which leads to each component of the coarse-grid operator being formed by Galerkin-coarsening the respective component in the fine-grid operator.

---

[3] https://www.gauss-centre.eu

The inverse of the sub-operator $Z \coloneqq BW^{-1}B^T$ is approximated as follows: assembling $Z$ would consume a prohibitively large amount of memory in case of a fine-grid operator, but $Z$ exists only on the pressure space, that is, due to the ($\mathbb{P}_1$-iso-$\mathbb{P}_2$, $\mathbb{P}_1$)-discretization, a coarser grid. Furthermore, $Z$ is only a scalar operator, holding another factor of $\frac{1}{9}$ (3D) fewer DoFs than the dominating part, the viscous operator $A_L$. Since $\text{diag}(A)^{-1}$ represents only a diagonal scaling, we do not see much fill-in in the triple-product. Thereby, $Z$ is essentially a sparse, two-levels-coarser operator and comparatively cheap to store, which we quantify in Section 5. Assembling enables us to solve $Z$ very rapidly using a HYPRE AMG-cycle [17] preconditioned CG, to a tolerance depending on the outer FGMRES tolerance. We switch from the default HYPRE configuration to three steps of aggressive coarsening paired with the HMIS-coarsening scheme. This is required to avoid excessive memory consumption in the AMG coarser-grid operators and declining scalability.

## REFERENCES

[1] M. ADAMS, M. BREZINA, J. HU, AND R. TUMINARO, *Parallel multigrid smoothing: Polynomial versus Gauss–Seidel*, Journal of Computational Physics, 188 (2003), pp. 593–610, https://doi.org/10.1016/S0021-9991(03)00194-3.

[2] R. E. ALCOUFFE, A. BRANDT, J. E. DENDY, JR., AND J. W. PAINTER, *The multi-grid method for the diffusion equation with strongly discontinuous coefficients*, SIAM Journal on Scientific and Statistical Computing, 2 (1981), pp. 430–454, https://doi.org/10.1137/0902035.

[3] D. ARNDT, N. FEHN, G. KANSCHAT, K. KORMANN, M. KRONBICHLER, P. MUNCH, W. A. WALL, AND J. WITTE, *ExaDG: High-order discontinuous galerkin for the exa-scale*, in Software for Exascale Computing - SPPEXA 2016-2019, H.-J. Bungartz, S. Reiz, B. Uekermann, P. Neumann, and W. E. Nagel, eds., vol. 136, Springer International Publishing, 2020, pp. 189–224, https://doi.org/10.1007/978-3-030-47956-5_8. Series Title: Lecture Notes in Computational Science and Engineering.

[4] S. BAUER, H.-P. BUNGE, D. DRZISGA, S. GHELICHKHAN, M. HUBER, N. KOHL, M. MOHR, U. RÜDE, D. THÖNNES, AND B. WOHLMUTH, *TerraNeo — Mantle Convection Beyond a Trillion Degrees of Freedom*, in Softw. Exascale Comput. - SPPEXA 2016-2019, H.-J. Bungartz, S. Reiz, B. Uekermann, P. Neumann, and W. Nagel, eds., vol. 136 of Lecture Notes in Computational Science and Engineering, Springer, 2020, pp. 569–610, https://doi.org/10.1007/978-3-030-47956-5_19.

[5] M. BERCOVIER AND O. PIRONNEAU, *Error estimates for finite element method solution of the stokes problem in the primitive variables*, Numerische Mathematik, 33 (1979), pp. 211–224, https://doi.org/10.1007/BF01399555.

[6] F. BÖHM, D. BAUER, N. KOHL, C. L. ALAPPAT, D. THÖNNES, M. MOHR, H. KÖSTLER, AND U. RÜDE, *Code generation and performance engineering for matrix-free finite element methods on hybrid tetrahedral grids*, SIAM Journal on Scientific Computing, 47 (2025), pp. B131–B159, https://doi.org/10.1137/24M1653756.

[7] C. BURSTEDDE, O. GHATTAS, M. GURNIS, T. ISAAC, G. STADLER, T. WARBURTON, AND L. WILCOX, *Extreme-scale amr*, in SC '10: Proceedings of the 2010 ACM/IEEE International Conference for High Performance Computing, Networking, Storage and Analysis, 2010, pp. 1–12, https://doi.org/10.1109/SC.2010.25.

[8] T. C. CLEVENGER AND T. HEISTER, *Comparison between algebraic and matrix-free geometric multigrid for a stokes problem on adaptive meshes with variable viscosity*, Numerical Linear Algebra with Applications, 28 (2021), p. e2375, https://doi.org/10.1002/nla.2375.

[9] T. C. CLEVENGER, T. HEISTER, G. KANSCHAT, AND M. KRONBICHLER, *A flexible, parallel, adaptive geometric multigrid method for fem*, ACM Trans. Math. Softw., 47 (2020), https://doi.org/10.1145/3425193.

[10] H. DE STERCK, U. M. YANG, AND J. J. HEYS, *Reducing complexity in parallel algebraic multigrid preconditioners*, SIAM Journal on Matrix Analysis and Applications, 27 (2006), pp. 1019–1039, https://doi.org/10.1137/040615729.

[11] Y. DEUBELBEISS AND B. KAUS, *Comparison of eulerian and lagrangian numerical techniques for the stokes equations in the presence of strongly varying viscosity*, Physics of the Earth and Planetary Interiors, 171 (2008), pp. 92–111, https://doi.org/10.1016/j.pepi.2008.06.023. Recent Advances in Computational Geodynamics: Theory, Numerics and Applications.

[12] C. R. DOHRMANN, *A preconditioner for substructuring based on constrained energy minimization*, SIAM Journal on Scientific Computing, 25 (2003), pp. 246–258, https://doi.org/10.1137/S1064827502412887.

[13] H. ELMAN, V. E. HOWLE, J. SHADID, R. SHUTTLEWORTH, AND R. TUMINARO, *Block preconditioners based on approximate commutators*, SIAM Journal on Scientific Computing, 27 (2006), pp. 1651–1668, https://doi.org/10.1137/040608817.

[14] H. C. ELMAN, D. J. SILVESTER, AND A. J. WATHEN, *Finite elements and fast iterative solvers: with applications in incompressible fluid dynamics*, Oxford university press, 2014.

[15] R. D. FALGOUT AND J. B. SCHRODER, *Non-galerkin coarse grids for algebraic multigrid*, SIAM Journal on Scientific Computing, 36 (2014), pp. C309–C334, https://doi.org/10.1137/130931539.

[16] R. D. FALGOUT AND J. B. SCHRODER, *Non-galerkin coarse grids for algebraic multigrid*, SIAM Journal on Scientific Computing, 36 (2014), pp. C309–C334, https://doi.org/10.1137/130931539, https://doi.org/10.1137/130931539, https://arxiv.org/abs/https://doi.org/10.1137/130931539.

[17] R. D. FALGOUT AND U. M. YANG, *hypre: A library of high performance preconditioners*, in Computational Science — ICCS 2002, P. M. A. Sloot, A. G. Hoekstra, C. J. K. Tan, and J. J. Dongarra, eds., 2002, pp. 632–641, https://doi.org/10.1007/3-540-47789-6_66.

[18] N. FEHN, P. MUNCH, W. A. WALL, AND M. KRONBICHLER, *Hybrid multigrid methods for high-order discontinuous galerkin discretizations*, Journal of Computational Physics, 415 (2020), p. 109538, https://doi.org/10.1016/j.jcp.2020.109538.

[19] B. GMEINER, M. HUBER, L. JOHN, U. RÜDE, AND B. WOHLMUTH, *A quantitative performance study for stokes solvers at the extreme scale*, Journal of Computational Science, 17 (2016), pp. 509–521, https://doi.org/10.1016/j.jocs.2016.06.006. Recent Advances in Parallel Techniques for Scientific Computing.

[20] M. GRIEBEL, B. METSCH, D. OELTZ, AND M. A. SCHWEITZER, *Coarse grid classification: a parallel coarsening scheme for algebraic multigrid methods*, Numerical Linear Algebra with Applications, 13 (2006), pp. 193–214, https://doi.org/10.1002/nla.482.

[21] X. HE, T. LIN, Y. LIN, ET AL., *Immersed finite element methods for elliptic interface problems with non-homogeneous jump conditions*, International Journal of numerical analysis and modeling, 8 (2011), pp. 284–301.

[22] A. KLAWONN, P. RADTKE, AND O. RHEINBACH, *Feti-dp methods with an adaptive coarse space*, SIAM Journal on Numerical Analysis, 53 (2015), pp. 297–320.

[23] N. KOHL, D. BAUER, F. BÖHM, AND U. RÜDE, *Fundamental data structures for matrix-free finite elements on hybrid tetrahedral grids*, Int. J. Parallel Emergent Distrib. Syst., 39 (2024), pp. 51–74, https://doi.org/10.1080/17445760.2023.2266875.

[24] N. KOHL AND U. RÜDE, *Textbook Efficiency: Massively Parallel Matrix-Free Multigrid for the Stokes System*, SIAM J. Sci. Comput., 44 (2022), pp. C124–C155, https://doi.org/10.1137/20M1376005.

[25] V. A. P. MAGRI, R. D. FALGOUT, AND U. M. YANG, *A new semistructured algebraic multigrid method*, SIAM Journal on Scientific Computing, 45 (2023), pp. S439–S460, https://doi.org/10.1137/21M1434118, https://doi.org/10.1137/21M1434118, https://arxiv.org/abs/https://doi.org/10.1137/21M1434118.

[26] D. J. RIXEN AND C. FARHAT, *A simple and efficient extension of a class of substructure based preconditioners to heterogeneous structural mechanics problems*, International Journal for Numerical Methods in Engineering, 44 (1999), pp. 489–516.

[27] J. RUDI, A. C. I. MALOSSI, T. ISAAC, G. STADLER, M. GURNIS, P. W. J. STAAR, Y. INEICHEN, C. BEKAS, A. CURIONI, AND O. GHATTAS, *An extreme-scale implicit solver for complex pdes: highly heterogeneous flow in earth's mantle*, in SC '15: Proceedings of the International Conference for High Performance Computing, Networking, Storage and Analysis, 2015, pp. 1–12, https://doi.org/10.1145/2807591.2807675.

[28] J. RUDI, G. STADLER, AND O. GHATTAS, *Weighted bfbt preconditioner for stokes flow problems with highly heterogeneous viscosity*, SIAM Journal on Scientific Computing, 39 (2017), pp. S272–S297, https://doi.org/10.1137/16M108450X, https://doi.org/10.1137/16M108450X, https://arxiv.org/abs/https://doi.org/10.1137/16M108450X.

[29] J. W. RUGE AND K. STÜBEN, *4. Algebraic Multigrid*, pp. 73–130, https://doi.org/10.1137/1.9781611971057.ch4.

[30] K. STÜBEN, *A review of algebraic multigrid*, in Numerical Analysis: Historical Developments in the 20th Century, C. Brezinski and L. Wuytack, eds., Elsevier, Amsterdam, 2001, pp. 331–359, https://doi.org/10.1016/B978-0-444-50617-7.50015-X.

[31] H. SUNDAR, G. BIROS, C. BURSTEDDE, J. RUDI, O. GHATTAS, AND G. STADLER, *Parallel geometric-algebraic multigrid on unstructured forests of octrees*, in SC '12: Proceedings of

the International Conference on High Performance Computing, Networking, Storage and Analysis, 2012, pp. 1–11, https://doi.org/10.1109/SC.2012.91.
[32] C. THIEULOT AND W. BANGERTH, *On the choice of finite element for applications in geodynamics*, Solid Earth, 13 (2022), pp. 229–249, https://doi.org/10.5194/se-13-229-2022.
[33] U. TROTTENBERG, C. W. OOSTERLEE, AND A. SCHULLER, *Multigrid methods*, Academic press, 2001.
[34] R. VERFÜRTH, *A posteriori error estimation techniques for finite element methods*, OUP Oxford, 2013.
[35] M. WEINZIERL AND T. WEINZIERL, *Quasi-matrix-free hybrid multigrid on dynamically adaptive cartesian grids*, ACM Trans. Math. Softw., 44 (2018), https://doi.org/10.1145/3165280.